\documentclass[10pt,conference]{IEEEtran}
\usepackage{multirow}
\usepackage{array}
\usepackage{makecell}
\usepackage{float}
\usepackage{hyperref}
\hypersetup{hidelinks}
\usepackage{url}
\usepackage[marginal]{footmisc}
\IEEEoverridecommandlockouts
\usepackage{cite}
\usepackage{amsmath,amssymb,amsfonts}
\usepackage{algorithmic}
\usepackage{graphicx}
\usepackage{textcomp}
\usepackage{xcolor}
\usepackage{balance}
\usepackage{booktabs}
\usepackage{enumerate}
\usepackage{enumitem}
\usepackage{threeparttable}
\def\BibTeX{{\rm B\kern-.05em{\sc i\kern-.025em b}\kern-.08em
    T\kern-.1667em\lower.7ex\hbox{E}\kern-.125emX}}
\begin{document}

\title{Understanding Architecture Erosion:\\The Practitioners' Perceptive}

\author{
    \IEEEauthorblockN{Ruiyin Li$^{1,2}$, Peng Liang$^{1*}$\thanks{\indent This work was partially funded by the National Key R\&D Program of China with Grant No. 2018YFB1402800. The authors gratefully acknowledge the financial support from the China Scholarship Council.}, Mohamed Soliman$^2$, Paris Avgeriou$^2$}
    \IEEEauthorblockA{$^1$ School of Computer Science, Wuhan University, Wuhan, China}
    \IEEEauthorblockA{$^2$ Department of Mathematics and Computing Science, University of Groningen, Groningen, The Netherlands}
    \IEEEauthorblockA{\{\href{mailto:ryli_cs@whu.edu.cn}{ryli\_cs}, \href{mailto:liangp@whu.edu.cn}{liangp}\}@whu.edu.cn, \{\href{mailto:m.a.m.soliman@rug.nl}{m.a.m.soliman}, \href{mailto:p.avgeriou@rug.nl}{p.avgeriou}\}\href{mailto:@rug.nl}{@rug.nl}}
}

\maketitle

\begin{abstract}
As software systems evolve, their architecture is meant to adapt accordingly by following the changes in requirements, the environment, and the implementation. However, in practice, the evolving system often deviates from the architecture, causing severe consequences to system maintenance and evolution. This phenomenon of architecture erosion has been studied extensively in research, but not yet been examined from the point of view of developers. In this exploratory study, we look into how developers perceive the notion of architecture erosion, its causes and consequences, as well as tools and practices to identify and control architecture erosion. To this end, we searched through several popular online developer communities for collecting data of discussions related to architecture erosion. Besides, we identified developers involved in these discussions and conducted a survey with 10 participants and held interviews with 4 participants. Our findings show that: (1) developers either focus on the structural manifestation of architecture erosion or on its effect on run-time qualities, maintenance and evolution; (2) alongside technical factors, architecture erosion is caused to a large extent by non-technical factors; (3) despite the lack of dedicated tools for detecting architecture erosion, developers usually identify erosion through a number of symptoms; and (4) there are effective measures that can help to alleviate the impact of architecture erosion.
\end{abstract}


\section{Introduction}\label{sec:Introduction}
Ideally, during the lifespan of a software system, its software architecture is constantly modified to satisfy new requirements and accommodate changes in the environment; therefore, the evolution of the architecture is aligned with the evolution of the software system \cite{RN211}. However, with the increasing complexity and changing requirements, the brittleness of a software system may increase, and the implementation may deviate from the architecture over time \cite{RN222}. This divergence between the intended and the implemented architecture is often called \textit{architecture erosion} \cite{RN221}.

Architecture Erosion (AEr) is not a new concept. Perry and Wolf \cite{RN222} explained around 30 years ago how a slowly eroding architecture makes it harder for new developers to understand the original system design. The phenomenon of AEr has been described using different terms in the literature, such as architectural decay \cite{RN210, RN213}, architecture degeneration \cite{RN214}, architecture degradation \cite{RN215}, and design erosion \cite{RN212}. AEr can significantly affect software development. Specifically, AEr can decrease software performance \cite{RN408}, substantially increase evolutionary costs \cite{RN409}, and degrade software quality \cite{RN364}, \cite{RN221}. Moreover, in an eroded architecture, code changes and refactorings could introduce new bugs and aggravate the brittleness of the system \cite{RN222}. Given such negative consequences, it is critical to investigate how far practitioners are aware of it and how they perceive it in their daily work.

Although AEr has been studied in the literature, there is little knowledge about the current state of practice of AEr from the perspective of developers. Hence, we attempted in this work an in-depth exploration of the viewpoints of developers on the notion of AEr, the causes and consequences of AEr, the used practices and tools for detecting AEr, and the measures to control and prevent AEr. To find developers with knowledge and experience on AEr, we looked into online developer communities, similar to the recent studies \cite{RN258}, \cite{RN171}, \cite{RN406}. These communities cover a wide spectrum of topics on software development, where practitioners share their development experience, ask questions and get responses. Specifically, we collected data about AEr from developer discussions in six popular online developer communities. In addition, we also leveraged the communities to reach out to developers who took part in the discussions, and thus were aware of and experienced on AEr. We then collected their opinions by conducting a mini survey with 10 participants and holding interviews with 4 participants. Finally, we analyzed the collected data from all the data sources (i.e., posts, questions and answers, surveys, interviews) using Constant Comparison \cite{RN433}.

The contributions of this work are threefold: (1) we explored how developers describe the phenomenon of AEr and how they perceive its manifestation; (2) we categorized and analyzed the diverse causes of AEr and the potential consequences to development from the developers' perspective; and (3) we investigated the effective tools and practices for detecting AEr, and presented the measures suggested by developers to control and prevent AEr.

The paper is organized as follows: Section \ref{sec:Related Work} introduces related work on AEr and online developer communities. Section \ref{sec:Study Design} elaborates on the study design. The results of each research question are presented in Section \ref{sec:results}, and their implications are discussed in Section \ref{sec:discussion}. Section \ref{sec:Threats} examines the threats to validity, while Section \ref{sec:Conclusions} summarizes this work and outlines the directions for future research.

\section{Related Work}\label{sec:Related Work}
\subsection{Architecture Erosion}\label{subsec:terms}
Several studies have explored the AEr phenomenon, which is described in various terms, such as ``\textit{architecture erosion}" \cite{RN221}, ``\textit{architecture decay}" \cite{RN302}, ``\textit{architecture degradation}” \cite{RN215}, ``\textit{software deterioration}” \cite{RN271}, ``\textit{architectural degeneration}” \cite{RN252}, and “\textit{software degradation}” \cite{RN272}.

Many studies focus on AEr with the purpose of identifying and repairing eroded architectures. For example, Wang \textit{et al.} \cite{RN329} proposed a multilevel method for detecting and repairing AEr based on architecture quality. Le \textit{et al.} \cite{RN213} identified AEr by analyzing architectural smells and their relationships between reported issues. 
De Silva and Balasubramaniam \cite{RN221} conducted a comprehensive survey on how to control AEr and the techniques for restoring and repairing eroded architectures, and they found that academic methods had limited adoption in industry.
While the aforementioned studies proposed approaches for detecting and repairing AEr, to the best of our knowledge, there are no studies that investigate AEr from the perspective of developers. Therefore, our study differs from the existing studies in that it offers an examination of the phenomenon from the developers' perspective, including its concept, the causes and consequences, as well as the tools and approaches used in practice to detect and control AEr.

\subsection{Online Developer Communities}\label{subsec:Communities}
There are millions of software practitioners who are active in online developer communities. They exchange knowledge, share experience, and provide an abundance of valuable discussions about software development. Such communities have been extensively utilized to conduct studies in the field of software engineering. For example, Stack Overflow, which is the largest and most visited Q\&A website, has been used as a source to efficiently identify architecturally relevant knowledge \cite{RN273}, investigate the relationships between architecture patterns and quality attributes \cite{RN274}, examine users’ behaviors and topic trends \cite{RN278}, \cite{RN279}, study architecture smells \cite{RN171}, and explore anti-patterns and code smells \cite{RN258}. In addition, other popular online developer communities are also used for research in software engineering, for example, understanding social and technical factors of contributions in GitHub \cite{RN280}.

Considering the advantages of online developer communities (such as the vast amount of available information, fast responses to questions, and diverse solutions to engineering problems), we decided to use them as our data source. In addition to mining data from these communities, we went one step further and contacted developers from these communities to collect more in-depth data. This allowed us to target subjects that have experience in the topic of AEr, while also to use two more sources (survey and interviews) for the sake of triangulation \cite{RN50} (see Section \ref{sec:Study Design}).

\section{Study Design}\label{sec:Study Design}
We formulated the goal of this study based on the Goal-Question-Metric approach \cite{RN117} as follows: \textbf{analyze} \textit{the perception of architecture erosion in practice} \textbf{for the purpose of} \textit{understanding} \textbf{with respect to} \textit{the notion, causes, consequences, detection and control of architecture erosion} \textbf{from the point of view of} \textit{developers} \textbf{in the context of} \textit{industrial software development}. In the following subsections, we explain the Research Questions (RQs), motivation, and the corresponding research steps and methods used to answer the RQs.

\subsection{Research Questions}\label{subsec:RQs}
\textbf{RQ1: How do developers describe architecture erosion in software development?}

\indent\indent\textbf{RQ1.1: Which terms do developers use to indicate architecture erosion?}

\indent\indent\textbf{RQ1.2: How does architecture erosion manifest according to developers?}

Researchers define AEr differently, such as architectural decay \cite{RN210}, \cite{RN213}, architecture degeneration \cite{RN214}, architecture degradation \cite{RN215} (see Section \ref{subsec:terms}). Through this RQ, we investigate what terms and aspects are used by developers (practitioners) to discuss the phenomenon of AEr and how they describe its manifestation in practice. This can provide insights on how practitioners communicate this phenomenon and how they describe it according to their own experience.

\textbf{RQ2: What are the causes and consequences of architecture erosion from the perspective of developers?}

There could be many factors leading to AEr, e.g., architecture smells \cite{RN213} and violations of architecture decisions \cite{RN245}. Furthermore, an eroded architecture comes with severe consequences to development, such as slowing down development or hampering maintenance. Through this RQ, we intend to identify the potential causes and consequences of AEr in practice. This can help to confirm whether the causes and consequences that have been reported in the literature hold in practice, and whether new ones come to light.

\textbf{RQ3: What practices and tools are used to detect architecture erosion in software development?}

Developers employ a number of practices and tools to identify potential architecture erosion in a system, or assess the quality of an eroded architecture. A list of such practices and tools used by practitioners can help other developers make informed decisions when dealing with AEr, and can inspire researchers to develop new approaches and tools.

\textbf{RQ4: What measures are employed to control architecture erosion in software development?}

After AEr is detected, it needs to be controlled to prevent it from hurting the system. By answering this RQ, we want to identify the measures taken by developers for addressing AEr in practice and the effect of these measures. Being aware of these measures has a direct benefit to practitioners, as the measures can help to prolong the lifetime of systems, and save substantial maintenance effort \cite{RN221}. Moreover, categorizing the measures against AEr can provide insights to researchers for refining and extending existing approaches (e.g., \cite{RN392}).

\subsection{Research Process}\label{subsec:Research Process}
To answer the RQs, we need to collect the opinions, knowledge, and experiences about AEr from practitioners. Considering that many practitioners may not be familiar with the concept and terms of AEr, we need an efficient mechanism to find practitioners who have the knowledge about AEr.

\begin{figure*}[htb]
	\centering
	\includegraphics[width=\linewidth]{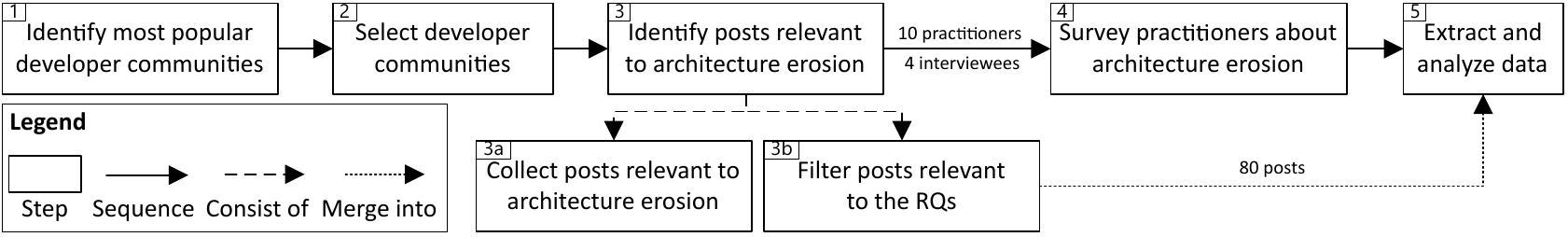}
	\caption{Overview of the research process}\label{F:1}
\end{figure*}

Online developer communities are regarded as an important platform for practitioners to communicate and share their knowledge and experience \cite{RN260}. For instance, Stack Overflow contains 20 million asked questions and 13 million users as of 21 January 2021. Hence, we decided to first collect data from the most popular developer communities and then contact the developers who were involved in the discussions about AEr. In detail, we followed five steps for data collection and analysis (see Figure \ref{F:1}), as elaborated in the following paragraphs.

\noindent\textbf{(1) Identify most popular developer communities}

There is no dedicated communities/websites for discussing issues of software architecture. Therefore, we planned to collect data from developer communities, and we first identified the most popular developer communities. We conducted a web search using Google in two steps:
\begin{enumerate}
\item \textit{Execute search queries}: We searched in Google using the search terms: ``popular/ranking/top/best online developer communities/forums”.
\item \textit{Identify websites}: We counted the frequency of websites mentioned in the search results, and ranked them to obtain the most frequently mentioned websites (see Table \ref{T:communities}). After the eighth most frequently mentioned websites, the rest had much lower frequencies so we considered the top eight. The complete list of websites and the corresponding frequencies are available online \cite{RN288}.
\end{enumerate}

\noindent\textbf{(2) Select developer communities}

We conducted a pilot search on each of the eight most popular developer communities (see Step (1) and Table \ref{T:communities}). For each developer community, we searched using the terms: ``architecture erosion”, ``architecture degradation”, ``architecture decay”. We then checked a sample from the top returned results, and determined their relevance to the topic of ``architectural erosion". Based on this pilot search, we selected six developer communities (see Table \ref{T:communities}), and excluded two (GitHub and Stack Exchange). The justification of excluding GitHub and Stack Exchange is presented below, and we also discussed the threat of this process in Section \ref{sec:Threats}.

\begin{itemize}[leftmargin=*]
\item When conducting the pilot study on GitHub, there are a large number of search results from \textit{Issues} and \textit{Commits} on GitHub, but a very low percentage of the results relate to AEr (less than 0.1\%). Therefore, the data extraction from GitHub would result in a very low return on investment.
\item Stack Exchange is a Q\&A website on topics in diverse fields, but not limited to software development. Stack Overflow is the largest sub-website of Stack Exchange and contains a number of discussion on software architecture \cite{RN569}. Thus, to avoid duplicated search results, we excluded Stack Exchange and retained Stack Overflow.
\end{itemize}

\begin{table}[t]
    \footnotesize
    \centering
    \caption{Eight most popular online developer communities}\label{T:communities}
    \begin{threeparttable}
    \begin{tabular}{|m{0.6mm}<{\centering}|m{18mm}|m{32.4mm}|m{4mm}<{\centering}|m{5mm}<{\centering}|m{5mm}<{\centering}|}
    \hline
    \# & \textbf{Website} & \textbf{URL Link} & \textbf{S}\tnote{1} & \textbf{\# R}\tnote{2} & \textbf{\# A}\tnote{3} \\\hline
    1 & Stack Overflow    & https://stackoverflow.com/             & Yes & 3973 & 39\\\hline
    2 & Reddit            & \multicolumn{1}{m{32.4mm}|}{\url{https://www.reddit.com/r/programming/}}   & Yes & 38 & 4\\\hline
    3 & Dzone             & https://dzone.com/                     & Yes & 556 & 26\\\hline
    4 & Hack News         & https://news.ycombinator.com/          & Yes & 625 & 8\\\hline
    5 & GitHub            & https://github.com/                    & No  & - & -\\\hline
    6 & Stack Exchange    & \multicolumn{1}{m{32.4mm}|}{\url{https://www.stackexchange.com/}}          & No  & - & -\\\hline
    7 & Code Project      & https://www.codeproject.com/            & Yes & 821 & 2\\\hline
    8 & Sitepoint         & \multicolumn{1}{m{32.4mm}|}{\url{https://www.sitepoint.com/community/}}   & Yes & 61 & 1\\\hline
    \multicolumn{3}{|c|}{Total} & 6 & 6074 & 80\\\hline
    \end{tabular}
    \begin{tablenotes}
        \footnotesize
        \item[1-3] S = Selected communities, R = Number of retrieved posts, A = Number of analyzed posts 
      \end{tablenotes}
      \end{threeparttable}
\end{table}

\noindent\textbf{(3) Identify posts relevant to architecture erosion}

In this step, we collected posts from the selected developer communities (see Step (2) and Table \ref{T:communities}), which discuss AEr and can provide answers to our RQs (see Section \ref{subsec:RQs}). Specifically, we performed two sub-steps: collecting (3a) and filtering posts (3b), as detailed below.


\noindent\textbf{(3a) Collect posts relevant to architecture erosion}

To identify the most suitable terms for capturing posts relevant to AEr, we experimented with several terms within the developer communities. After a pilot search with several combinations of search terms, we decided to use the search query below: (additional details of the pilot search with other search terms are available online \cite{RN288}).

\textit{((architecture OR architectural OR structure OR structural) AND (decay OR erode OR erosion OR degrade OR degradation OR deteriorate OR deterioration))}

To execute the search query effectively, we used different search engines for each developer community (see Table \ref{T:communities}), depending on the effectiveness of internal search engines offered by each developer community. For Stack Overflow, Sitepoint, and Hack News, we used their internal search engines. In contrast, Dzone, Reddit, and Code Project lack an effective search engine, and thus we used Google.

For each developer community, we executed the same search query, and removed duplicated search results. We executed the search manually for all the developer communities except for Stack Overflow, for which we built a crawler to automatically execute queries, collect and remove duplicated posts. We identified duplicated Stack Overflow posts by comparing the title of each post with other posts; using the title results in less duplicates, as posts may have different IDs but contain the same information (e.g., Stack Overflow posts with IDs ``\textit{36475408}” and ``\textit{36475800}”\footnote{To access the post on Stack Overflow, add the ID number to the URL: https://stackoverflow.com/questions/ID}).


\noindent\textbf{(3b) Filter posts relevant to the RQs}

We manually checked the collected posts to ensure their relevance to answering the RQs (see Section \ref{subsec:RQs}). To achieve this, we specified the following inclusion and exclusion criteria (Boolean AND is used to connect the inclusion criteria):

\textbf{1) Inclusion criteria}
\begin{itemize}[leftmargin=*]
\item \textbf{I1}: The post (and its answers) discusses architecture erosion in software development.
\item \textbf{I2}: The post (and its answers) is relevant to answering the RQs.
\end{itemize}

\textbf{2) Exclusion criteria}
\begin{itemize}[leftmargin=*]
\item \textbf{E1}: When the content of two or more posts contain similar information, we exclude the one less relevant to the RQs.
\end{itemize}

Before the formal data filtering through manual inspection, to reach an agreement about the criteria, we conducted a pilot filtering step with 50 posts randomly-selected from the 3,973 retrieved posts on Stack Overflow, which were checked by the first and second author independently. We then calculated the Cohen’s Kappa coefficient \cite{RN261} of the pilot filtering results to measure the inter-rater agreement between the two authors; this achieved an agreement of 0.728. Any disagreements and uncertain judgments on the posts were discussed between the two authors until a consensus was reached. Then the first author conducted the formal filtering with all the retrieved posts (see Table \ref{T:communities}), and discussed with the second author any uncertainties in order to get a consensus. Eventually, we collected in total 80 posts related to our RQs (see Table \ref{T:communities}). Note that, 3,973 posts were originally retrieved from Stack Overflow, but we excluded a large number of irrelevant posts (e.g., topics about image erosion, array decay/degradation, learning rate decay).

\noindent\textbf{(4) Survey practitioners about architecture erosion}

While the data from community posts are valuable, we wanted to enrich them with other data sources; this helps also to achieve data source triangulation \cite{RN50}. Thus we conducted a survey and interviews for collecting more data from another two sources (i.e., questionnaires and interviews). The prerequisite of conducting surveys and interviews is to find qualified participants, and collecting posts that discuss AEr (as explained in Step (3)) gives us the chance to identify the practitioners, who are potentially knowledgeable about AEr. These practitioners who were involved in the discussion of the collected posts can provide credible answers to the RQs.

To identify practitioners, we manually inspected the profiles of users, who were involved in the discussion of the 80 collected and filtered posts (from Step (3)). However, some users provided minimal information in their profile about their identity. This prevented us from identifying all potential practitioners. After inspecting all user profiles, we found 38 valid user profiles, from which we could discern their identity and contact information. We further searched for these 38 practitioners on LinkedIn, in order to validate their identity, and gather missing background and contact information. We note that the anonymity of these developers is preserved in both the manuscript and the accompanying dataset \cite{RN288}.

We contacted the 38 practitioners, and sent each practitioner two invitations. The first one was for answering the survey and the second invited them for participating in a one-to-one interview. 13 out of the 38 practitioners (34.2\%) accepted our invitations for either filling out the survey (10 practitioners) or conducting a one-to-one interview (4 practitioners); note that one developer took part in both the survey and the interview. We got a response rate of 34.2\%, which is much higher than the general response rate in empirical software engineering research around 5\% \cite{RN116}. Table \ref{T:Background} provides the background information about the 13 practitioners.

\begin{table}[t]
    \footnotesize
    \centering
    \caption{Background of the participants}\label{T:Background}
    \begin{threeparttable}
    \begin{tabular}{|c|c|c|c|c|m{27mm}|m{9mm}<{\centering}|}
        \hline
        \textbf{S/I}\tnote{*} & \textbf{EB}\tnote{1} & \textbf{YE}\tnote{2} & \textbf{YA}\tnote{3} & \textbf{YD}\tnote{4} & \textbf{Role} & \textbf{Location}\\\hline
        S & PhD & 44 &  28  &  -   & Director                     & UK\\\hline
        S & MSc & 15 &  -   & 10.3 & Senior Consultant            & Norway\\\hline
        S & BSc & 25 &  -   & 25   & Senior Software Engineer     & UK\\\hline
        S & PhD & 20 &  6.9 & 6.2  & Senior Lead Developer        & UK\\\hline
        S & MSc & 20 &  2   & 14.8 & Architect                    & Spain\\\hline
        S & PhD & 54 & 15.8 & 6.3  & Architecture Consultant      & Denmark\\\hline
        S & MSc & 16 &  -   & 15.3 & DevOps Coach                 & Norway\\\hline
        I & MSc & 27 &  16  & -    & CEO                          & USA\\\hline
        I & MSc & 20 &  2.1 & 15   & Lead Software Developer      & USA\\\hline
        B & MSc & 21 &  15  & 6    & Software Intelligence Expert & USA\\\hline
        S & PhD & 21 &  7.7 & 3.9  & CEO                          & USA\\\hline
        S & PhD & 16 &  -   & 16   & Research Scientist           & USA\\\hline
        I & MSc & 8  & 2.8  & 5.2  & Senior Software Engineer     & India\\\hline
    \end{tabular}
     \begin{tablenotes}
        \footnotesize
        \item[*] S = Survey, I = Interview, B = Both
        \item[1-4] EB = educational background, YE = years of experience in software development, YA = years of experience as architect, YD = years of experience as developer 
      \end{tablenotes}
      \end{threeparttable}
\end{table}

Based on the selected posts and our RQs, we used similar questions for both the customized surveys and interviews (available online \cite{RN288}). The questions were designed and refined iteratively by the authors. We decided to use open-ended questions to allow practitioners to express their own opinions; this is in line with our research goal to determine the practitioners' point of view on AEr. Before sending the survey to practitioners and conducting the interviews, we conducted a pilot survey with 2 developers who had the knowledge about AEr and helped to improve the clarity of the questions. Note that the pilot survey answers were not included in the data for analysis. The survey and interviews are detailed below.

\textbf{\textit{Survey}}: We sent the questionnaire to the 38 practitioners, and 10 of them accepted our invitation to fill in the questionnaire. To make the questionnaire more relevant to the practitioners and improve the response rate, we sent each practitioner a customized version of the questionnaire \cite{RN288}, which was aligned with the post that he/she was involved by using the same term of AEr as the practitioner used to prevent misunderstanding. The survey responses were collected by Google Forms and saved in a Word document for later extraction and analysis (see Step (5)).

\textbf{\textit{Interviews}}: We conducted semi-structured interviews by following the guidelines in software engineering \cite{RN405} with 4 practitioners, who accepted our invitation. The interviews were conducted by the third author (a postdoctoral fellow), and transcribed and analysed by the first and second authors (see Step (5)). For the interviews, we used similar questions with the survey as a basis. As is usual with semi-structured interviews, additional questions came up during the interview depending on the answers to specific questions; in such cases, we allowed the practitioners to freely express their opinions.

\noindent\textbf{(5) Data extraction and analysis}

We used the data items (see Table \ref{T:items}) to collect data from three sources, i.e., online developer communities (80 posts), the survey (10 answers), and interviews (4 interviews). 
The first author extracted the data and the second author subsequently reviewed it. To alleviate personal bias, any uncertainties and ambiguities in the extraction results were discussed between the first and second authors to reach an agreement.


Regarding data analysis, we used descriptive statistics (for data item D1 and D4) and Constant Comparison \cite{RN433} (for data items D1-D5) to analyze and categorize the extracted qualitative data. Constant Comparison can be used to generate concepts, categories, and theories through a systematic analysis of the qualitative data. For the five data items, the first author coded the data (around 400 annotations) using Constant Comparison and the second author checked the coding results during the two coding activities (i.e., open coding and selective coding); any divergence in the coding and categorization results were further discussed until the two authors reached an agreement. To effectively code and categorize data, we employed the MAXQDA tool\cite{RN112} to help conduct manual data coding
, which is a commercial tool for qualitative data analysis. All the data of this study is available online \cite{RN288}.

\begin{table}[t]
    \footnotesize
    \centering
    \caption{Mapping between the extracted data items and RQs}\label{T:items}
    \begin{tabular}{|c|m{23mm}|m{39mm}|c|}
        \hline
        \#&\textbf{Data Item}&\textbf{Description}&\textbf{RQ}\\\hline
        D1 & Description of AEr & Terms used to describe AEr and how it manifests & RQ1 \\\hline
        D2 & Causes of AEr & Potential factors leading to AEr & RQ2 \\\hline
        D3 & Consequences of AEr & Impacts and ramifications of AEr on development & RQ2 \\\hline
        D4 & Practices and tools for detecting AEr & Approaches used in practice to identify AEr & RQ3 \\\hline
        D5 & Measures to control AEr & Ways to address or manage the impact of AEr & RQ4 \\\hline
    \end{tabular}
\end{table}

\section{Results}\label{sec:results}
\subsection{RQ1 - Description of architecture erosion}\label{subsec:RQ1}
To answer RQ1.1, we first used descriptive statistics to analyze the frequencies of the used terms. We then categorized the terms used to describe AEr into five types (see Table \ref{T:terms}). We can see that most developers prefer to use \textit{erode/erosion} to describe AEr. The term \textit{degrade/degradation} comes second, followed closely by \textit{decay}. Another popular term, though not as popular as the first three, is \textit{deteriorate/deterioration}. In addition, we found several terms not included in our search terms are used to represent the AEr phenomenon when we manually checked the extracted data, such as \textit{software rot}, \textit{software entropy}, \textit{software aging}, and \textit{fundamental design flaw}. Note that, some developers used ambiguous or high level terms to describe AEr, like \textit{system degradation}; but it is not clear whether a term like  \textit{system degradation} refers to performance degradation or structure degradation of the system. For this reason, we excluded such vague terms.

\begin{table}[t]
    \footnotesize
    \centering
    \caption{Terms that developers used to describe architecture erosion}\label{T:terms}
    \begin{tabular}[width=\linewidth]{|m{15mm}<{\centering}|m{50mm}|m{2.5mm}<{\centering}|m{2.5mm}|}
    \hline
    \textbf{Type} & \textbf{Term} & \multicolumn{2}{c|}{\textbf{Count}}\\\hline
    \multirow[m]{5}*{\makecell{Erode\\/erosion}} & Architecture/architectural erode /erosion & 20 & \multirow[m]{5}*{26}\\\cline{2-3}
    ~ & Structure/structural erosion & 3 & ~ \\\cline{2-3}
    ~ & Software erosion & 1 & ~\\\cline{2-3}
    ~ & Project erosion & 1 & ~ \\\cline{2-3}
    ~ & Component erosion & 1 & ~\\\hline
    \multirow{6}*{\makecell{Degrade\\/degradation}} & Architecture/architectural degrade /degradation & 6&\multirow[m]{6}*{16}\\\cline{2-3}
    ~ & \makecell[l]{Structure/structural degrade /degradation} & 2 &~\\\cline{2-3}
    ~ & Project/product degrade /degradation& 3 & ~\\\cline{2-3}
    ~ & Code degrade & 1 & ~ \\\cline{2-3}
    ~  & Design degrade & 1 & ~ \\\cline{2-3}
    ~ & Module degrade & 1 & ~ \\\hline
    \multirow{6}*{Decay} & Architecture/architectural decay & 8 & \multirow[m]{6}*{15}\\\cline{2-3}
    ~ & Project decay & 2 & ~ \\\cline{2-3}
    ~ & Software decay & 2 & ~\\\cline{2-3}
    ~ & Design decay & 1 & ~\\\cline{2-3}
    ~ & Package structure decay & 1 & ~\\\cline{2-3}
    ~ & Structural decay & 1 & ~ \\\hline
    \multirow{3}*{\makecell{Deteriorate\\/deterioration}} & Architecture/architectural deteriorate /deterioration & 4 & \multirow[m]{3}*{8}\\\cline{2-3}
    ~ & Structure deterioration & 2 & ~\\\cline{2-3}
    ~ & Code quality deterioration & 2 &~\\\hline
    \multirow[m]{4}*{Others} & Software / design rot& 5 & \multirow[m]{4}*{9}\\\cline{2-3}
    ~ & Software entropy & 2 &~\\\cline{2-3}
    ~ & Software aging & 1 & ~\\\cline{2-3}
    ~ & Fundamental design flaw & 1 & ~\\\hline
    \end{tabular}
\end{table}

To answer RQ1.2, we analyzed the context of AEr mentioned by developers and categorized the manifestation of this phenomenon in four perspectives with Constant Comparison.

\textbf{From the \textit{structure} perspective}: the structure of an eroded architecture deviates from the intended architecture.
As mentioned by interviewee \#1, ``\textit{Architecture (erosion) is about the original architecture blueprint got lost when you can't know architecture structure and boundaries anymore}". A number of striking examples were reported. Consider, for example, the violation of design rules about encapsulation that breaks implemented abstract layers and can have long-lasting impact on maintenance (in the interest of performance). A similar problem is the accumulation of cyclic dependencies and increased coupling, both resulting in binding elements together that were intentionally separated by architects. Finally, other mentioned structural manifestations include dead/overlapping code, and obsolete or incompatible third-party libraries.

\textbf{From the \textit{quality} perspective}: an eroded architecture may not meet the original or current non-functional requirements; thus the system quality attributes are degraded. As developer \#1 stated ``\textit{to make changes error prone, or to no longer meet cross-functional requirements (performance, security, scalability, etc.)}". Our data indicates a negative effect of AEr mostly on reliability (mainly due to increasing error-proneness), performance, and user experience.

\textbf{From the \textit{maintenance} perspective}: an eroded architecture could be harder to understand, fix bugs, and refactor. As developer \#2 stated ``\textit{it is often very hard to understand the existing architecture, determine the extent of architectural decay, and identify architectural smells and metric violations}". Our data indicates that increasing complexity and technical debt often become common in eroded architectures, making bug-fixing and refactoring increasingly difficult.

\textbf{From the \textit{evolution} perspective}: an eroded architecture makes it hard or even impossible to plan the next evolution steps, e.g., which features to implement next or which technologies to adopt. As developer \#3 pointed out ``\textit{the unfortunate side effect of this [erosion] is that it becomes more and more difficult to add new visible features without breaking something}". Developer \#3 also commented that, when an architecture erodes over time, ``\textit{the stakeholders will notice instability, high maintenance cost and ridiculously high cost for adding or changing features}".

\subsection{RQ2 - Causes and consequences of architecture erosion}\label{subsec:RQ2}
\noindent\textbf{(1) Causes of architecture erosion}

In Table \ref{T:causes}, we list the potential causes leading to AEr, categorized into 12 types. \textbf{Inappropriate architecture changes} are the most frequently mentioned reason incurring AEr, and often happen in the maintenance and evolution phases in a number of ways: introducing new anomalies (e.g., cyclic dependencies), breaking architectural rules, introducing new architectural principles that are incompatible with existing frequent modifications leading to the accumulation of cyclic dependencies. It is hard to accurately anticipate the side effects of those architectural changes, as developers cannot fully understand which part of the functionality will be implicated. The inappropriate changes to an architecture increase the risk of undermining the architectural integrity (e.g., breaking the encapsulation rules), introducing new bugs and causing architecture smells (e.g., increasing superfluous dependencies).

Unlike inappropriate architecture changes that happen during maintenance, \textbf{architecture design defects} often occur in the design phase. It could be regarded as a hidden danger to sustainability, as they often cannot be discovered via static analysis. These flaws in the system architecture, eventually lead to a gap between the intended architecture and the implementation. For example, if the original system has a ``bad encapsulation", the implementation is likely to gradually deviate from the intended architecture as dependencies are created with the poorly encapsulated functionality over time. Developer \#16 commented that, AEr happens as ``\textit{the inherent flaws present in every initial design begin to surface}". Developer \#22 mentioned that ``\textit{if your core system abstractions are not clean, then the system is destined to degrade}''.

\textbf{Lack of management skills} is another common cause of AEr and examples include: assigning incompetent developers to do a job they do not fit, having unreasonable rewarding and punishment metrics in place (that could lead to a high turnover of staff), lacking proper training and education for developers (resulting, for example, in non-uniform coding standards), or lacking long-term strategies for architecture evolution. As developer \#15 stated ``\textit{some worse or mediocre developers who just are too uncomfortable with creative development and technical decision making are "promoted" to the management. As a result, everyone suffers, ..., it often creates disaster}".

The accumulation of \textbf{technical debt} is also a major cause leading to AEr. Technical debt \cite{RN404} is a short-term solution that may expedite development, but it might violate architectural principles, hampering refactoring and maintenance in general, and eventually deviating from the target architecture. As developer \#4 stated ``\textit{the short term benefit of finishing your task now by taking short cuts versus the long term risk of making the code less understandable}".

\textbf{Disconnection between architects and developers} manifests through three scenarios: (1) architects do not adequately monitor the implementation process; (2) developers do not actively participate in the architecting process; or (3) there is complete absence of architect roles. Architects need to guide and monitor the implementation of architecture design and developers also need to understand the rationale of architectural decisions. As mentioned by developer \#5 ``\textit{the software changes need special attention (architectural assessment) from software architects. If this does not happen, the architecture could erode or become overly complex}".

\textbf{Knowledge vaporization} is mostly due to developer turnover, and poorly documented architectural knowledge. It is a potential threat to project management. For instance, knowledge is lost when developers leave the team, while new developers may violate architecture design principles due to the lack of knowledge about the current architecture.

\textbf{Requirements changes} challenge the architecture sustainability, e.g., when the architecture is incompatible with the newly-added or changed requirements, developers need to remove some parts from the architecture and add extra ``patches"; this often impacts the maintainability and extensibility of the architecture. Note that, this kind of requirements are usually unforeseen and/or unplanned requirements (e.g., increasing demands for storage), that are in conflict with existing design rules and constraints.

\textbf{Lack of communication} has many obvious disadvantages to software development and particularly maintenance. For example, as mentioned by developer \#6 ``\textit{if some developers isolate themselves from others, this may reduce the communication complexity at the cost of increasing the program complexity}"; consequently this increased complexity makes it harder to implement the intended architecture correctly. Interviewee \#2 stated ``\textit{communication is key, for everyone on a team, including the architects and builders}". 

In many cases, due to quick iterations and releases, developers might ignore long-term architectural strategies. Particularly, \textbf{agile development} is considered as a cause of rapid AEr. This is a common and recurring issue faced by agile development teams, as described by developer \#7 ``\textit{No architecture will stay intact in the face of agile, evolving requirements}”. Recent studies also show that agile process may not make a project agile \cite{RN266} and architecture might become the bottleneck of agile projects. Additionally, \textbf{increasing complexity} could slowly degrade the architecture, make codebases less understandable, and gradually make it harder and harder to maintain and evolve the system.

\textbf{Lack of maintenance} is regarded as another cause of AEr. If the architectural components are outdated and maintainers do not constantly refactor and maintain the codebase to keep it tidy and clean (e.g., replacing obsolete third party libraries), the architecture is destined to erode. Finally, there are four less frequently mentioned causes of AEr, including environment change, business process change, business pressure, and treating quality concerns as second-class citizens. For example, when a business process changes, the architecture might become incompatible with the new process (i.e., erosion).


\begin{table}[b]
    \footnotesize
    \centering
    \caption{Causes of architecture erosion}\label{T:causes}
    \begin{tabular}[width=\linewidth]{|c<{\centering}|m{40mm}|m{17mm}|m{7mm}<{\centering}|}
    \hline
    \textbf{No.} & \textbf{Cause} & \textbf{Type} & \textbf{Count} \\\hline
    1 & Inappropriate architecture changes & Technical & 22 \\\hline
    2 & Architecture design defects & Technical & 15 \\\hline
    3 & Lack of management skills & Non-technical & 13 \\\hline
    4 & Technical debt & Technical & 11 \\\hline
    5 & \multicolumn{1}{m{40mm}|}{Disconnection between architects and developers} & Non-technical & 10 \\\hline
    6 & Knowledge vaporization & Non-technical & 9 \\\hline
    7 & Requirements change & Both & 9 \\\hline
    8 & Lack of communication & Non-technical & 8 \\\hline
    9 & Agile development & Technical & 8 \\\hline
    10 & Increasing complexity & Technical & 7\\\hline
    11 & Lack of maintenance & Both & 6\\\hline
    12 & Others (environment change, business process change, business pressure, quality concerns as 2nd-class) & Non-technical & 9 \\\hline
    \end{tabular}
\end{table}

\begin{table}[t]
    \footnotesize
    \centering
    \caption{Consequences of architecture erosion}\label{T:consequences}
    \begin{tabular}[width=\linewidth]{|c<{\centering}|m{56mm}|c|}
    \hline
    \textbf{No.} & \textbf{Consequence} & \textbf{Count} \\\hline
    1 & Hard to understand and maintain & 20 \\\hline
    2 & Run-time quality degradation & 13 \\\hline
    3 & Enormous cost to refactor & 11 \\\hline
    4 & Big ball of mud & 9 \\\hline
    5 & Slowing down development & 5\\\hline
    6 & High turnover rate & 3 \\\hline
    7 & Overall complexity & 2\\\hline
    \end{tabular}
\end{table}

\noindent\textbf{(2) Consequences of architecture erosion}

The consequences of AEr are presented in Table \ref{T:consequences}. \textbf{Hard to understand and maintain} is the most frequently mentioned consequence. For example, an eroded architecture with implicit dependencies might be difficult to maintain without breaking some dependencies, and developers may not understand the ramification of breaking these dependencies. \textbf{Run-time quality degradation} is another major consequence of AEr. Users perceive a compromise in run-time qualities, such as performance, reliability, and user experience.

Eroded architectures may incur an \textbf{enormous cost to refactor}. As developer \#8 said ``\textit{there is a risk that the refactoring cost is significant, possibly even too high to contemplate, resulting in a system is ``stuck" in an undesirable form}". Furthermore, due to the increasing complexity of systems (e.g., cyclic dependencies), \textbf{slowing down development} can be common during the development and maintenance phase. For example, time-to-delivery is delayed, while implementing new features and debugging can become extremely slow or even stagnant. In the worst case, an eroded architecture may render the system a \textbf{big ball of mud} \cite{RN421}, i.e., the system lacks a perceivable architecture. As developer \#9 stated ``\textit{As with all big ball of muds, the issue doesn't usually make itself apparent until there's some maintenance/enhancement needed}".

An eroded architecture can cause \textbf{high turnover rate}, as developers are forced to work on a messy architecture. Developer \#10 explained how this affects an organization: ``\textit{it's not the software that rots but instead the users and/or organization that decays}". 
Developer \#11 discussed the similarity between the software and the organization: ``\textit{Conway's Law would suggest that software architecture mirrors organization structure, so it too would become more brittle}". The high turnover further aggravates AEr, due to losing knowledge about system requirements and design decisions.
Additionally, the \textbf{overall complexity} of the architecture can increase drastically. Developer \#12 stated 
``\textit{architectural erosion starts to happen as you add capabilities and slowly increase software complexity}". The accumulation of complexity, if left un-controlled, bears the risk of bringing the entire project to a halt.

\subsection{RQ3 - Identifying architecture erosion}\label{subsec:RQ3}
\noindent\textbf{(1) Tools}

To answer RQ3, we collected the practices and tools employed to identify AEr. Table \ref{T:tools} lists the 13 tools collected and ranked according to their frequency mentioned by developers. We note that these tools practically identify issues in the architecture that can be considered as symptoms of AEr; in other words, none of the tools claims to specifically identify AEr per se. For example, Lattix \cite{RN309} is a commercial tool that allows users to create dependency models to identify architecture issues; NDepend \cite{RN309} can help users to find out architectural anomalies; Structure101 \cite{RN310} offers views of code organization and helps practitioners to better understand the code structure and dependencies for checking architecture conformance. Some tools are language-specific, such as JDepend \cite{RN313} and Archie (for Java), Designite (for C\#), while other tools support multiple programming languages.

\begin{table}[b]
    \footnotesize
    \centering
    \caption{Tools used to detect architecture erosion}\label{T:tools}
    \begin{tabular}{|m{3mm}<{\centering}|m{18mm}|m{43mm}|m{7mm}<{\centering}|}
    \hline
    \textbf{No.} & \textbf{Tool} & \textbf{Link} &\textbf{Count}  \\\hline
    1 & Lattix & \multicolumn{1}{m{43mm}|}{\url{https://www.lattix.com}} & 10 \\\hline
    2 & NDepend & \multicolumn{1}{m{43mm}|}{\url{https://www.ndepend.com}} & 10 \\\hline
    3 & Sonargraph & \multicolumn{1}{m{43mm}|}{\url{http://www.hello2morrow.com/products/sonargraph}} & 5 \\\hline
    4 & Structure101 & \multicolumn{1}{m{43mm}|}{\url{https://structure101.com/products/workspace}} & 4 \\\hline
    5 & Architecture-Quality-Evolution & \multicolumn{1}{m{43mm}|}{\url{https://github.com/tushartushar/ArchitectureQualityEvolution}} & 4 \\\hline
    6 & JDepend & \multicolumn{1}{m{43mm}|}{\url{https://github.com/clarkware/jdepend}} & 3 \\\hline
    7 & Designite & \multicolumn{1}{m{43mm}|}{\url{https://www.designite-tools.com/}} & 3 \\\hline
    8 & SonarQube & \multicolumn{1}{m{43mm}|}{\url{https://www.sonarqube.org}} & 2 \\\hline
    9 & SonarLint &\multicolumn{1}{m{43mm}|}{\url{https://www.sonarlint.org/}} & 1 \\\hline
    10 & Archie &\multicolumn{1}{m{43mm}|}{\url{https://github.com/ArchieProject/Archie-Smart-IDE}} & 1 \\\hline
    11 & Glasnostic &\multicolumn{1}{m{43mm}|}{\url{https://glasnostic.com/}} & 1 \\\hline
    12 & CodeScene &\multicolumn{1}{m{43mm}|}{\url{https://codescene.io/}} & 1 \\\hline
    13 & CAST & \multicolumn{1}{m{43mm}|}{\url{https://www.castsoftware.com/}} & 1 \\\hline
    \end{tabular}
\end{table}

\noindent\textbf{(2) Practices}

Apart from the tools mentioned by developers, we also found several general practices applied to identify AEr, as elaborated in the following paragraphs.

\textbf{Dependency Structure Matrix (DSM)} \cite{RN301} can be used to visually represent a system in the form of a square matrix. Developer \#13 mentioned ``\textit{DSMs are also a powerful way of setting and visualizing design rules. They make it easy to pinpoint violations to design rules}", which are typical symptoms of AEr. DSMs can visualize dependency relationships between packages (e.g., \cite{RN302}, \cite{RN303}); understanding such dependencies helps detect AEr during the maintenance phase.

\textbf{Software Composition Analysis (SCA)} \cite{RN305}, \cite{RN306} refers to the process that provides visibility of open source components in a system. As stated by developer \#14, ``\textit{Managing a product against software decay can be a nightmare, but again a good SCA tool should be able to take care of that. It should update the developers when a new open-source library becomes available}". Open source components are used in software products across all industries, and SCA can especially help to determine latent obsolete components that can typically cause AEr. There are also some SCA tools available, such as WhiteSource, Snyk, and Sonatype.

\textbf{Architecture Conformance Checking (ACC)} refers to the type of conformance between the implemented architecture and the intended architecture. The detection happens by identifying architectural violations in the implementation. As developer \#3 mentioned ``\textit{It is easier to pass a system from a development team to a maintenance team, especially when changes can automatically be checked for architectural conformance. Also new team members need less time to become productive because the code is easier to understand}".

\textbf{Architecture monitoring} refers to using tools to monitor the health of the architecture by detecting architecture issues. This is achieved by various techniques (e.g., Reflexion models \cite{RN415}) and metrics, such as coupling, size of files, hotspots \cite{RN264}). As developer \#13 mentioned ``\textit{When design rules are monitored, tight scheduling does not erode the architecture and, if it does, the consequences of time pressure can be tracked (architectural technical debt) and monitored}".

\textbf{Code review} is a continuous and systematic process conducted by developers or architects to identify mistakes in code, such as violations of design patterns. As developer \#1 mentioned ``\textit{when the team is small enough, using code reviews will be effective enough to prevent architectural erosion}".

\textbf{Checking the change of architectural smell density} is a method employed to detect AEr following the release timeline by statically comparing architectural smells in various versions. As developer \#17 stated ``\textit{Comparing absolute values of detected architecture smell instances across versions is not a good idea because the size of the code is also changing. Therefore, we compute architectural smell density. It is a normalized metric capturing number of smells per one thousand LOC}”. By observing the change rate of architectural smell density \cite{RN445}, developers can find out from which version, the architecture started to deteriorate.

\textbf{Architecture visualization} \cite{RN416} aims at representing architectural models and architectural design decisions using various visual notations. It helps to better understand the architecture and its evolution by visualizing the structure, metrics, and dependencies between architecture elements (e.g, components) in a project. Understanding architectural dependencies of a system through visualization is significant to detect the proliferation of violations \cite{RN415} and further erosion.

\subsection{RQ4 - Addressing architecture erosion}\label{subsec:RQ4}
To answer RQ4, we categorized the measures used to control AEr, as discussed in the following paragraphs.

\textbf{Architecture assessment} refers to the assessment process throughout the life cycle during architecture design (e.g., discovering and addressing the shortcomings of design decisions), during architecture implementation (e.g., monitoring and repairing possible violations of design rules), and during architecture evolution (e.g., choosing appropriate refactoring patterns to fix issues from new requirements or evaluating the risk of architectural changes). We note that architecture assessment is meant to be followed up with concrete actions; in other words, it is a measure that connects detecting and addressing AEr, as
developer \#5 mentioned ``\textit{Architecture erosion can happen in any software project where the architectural assessments are not part of the development process}".

\textbf{Periodic maintenance} refers to regular activities (e.g., code refactoring, bug-fixing, testing) aimed at keeping a system ``clean" and running smoothly. Developer \#18 stated ``\textit{you can test the architecture regularly every time you make changes to the code. This eliminates the worry about architectural erosion in your software}". Another developer \#19 urged ``\textit{don't leave `broken windows' un-repaired. Fix each one as soon as it is discovered}" (examples of ``broken windows" are bad designs, wrong decisions, or poor code). If there is insufficient time to conduct maintenance work right away, developers can create a list of pending problems and technical debt, and find a suitable time to pay off the debt and replace the temporary solutions.

\textbf{Architecture simplification}. When architectural complexity proliferates towards being uncontrollable, simplifying the architecture, and deliberately controlling the system size and complexity could be an option worthy of consideration. Developer \#12 mentioned ``\textit{There needs to be a continuous effort to simplify (refactor) the code. If not, architectural erosion starts to happen as you add capabilities and slowly increase software complexity}". Several developers mentioned migrating to a microservices architecture, as one prominent way to achieve this. Decomposing the original monolithic architecture into many small microservices can, to some extent, improve architectural extensibility and increase its resilience to AEr. 

\textbf{Architecture restructuring} is a drastic, yet effective means to control AEr. It is a much more pervasive change that concerns a large part of the architecture, compared to \textit{architecture simplification}, that merely tries to reduce complexity. Developer \#19 stated ``\textit{Sometimes, the best solution is simply to rewrite the application catering to the new requirements. But this is normally the worst case scenario. The cumbersome solution is to stop all new development, start by writing a set of tests and then redesign and rearchitect the whole solution}”. However, restructuring the original architecture to satisfy new requirements and keeping the system running smoothly, may require enormous time and effort, considering the famous example of Mozilla web browser \cite{RN431}.

\textbf{Organization optimization}. Hiring more capable team members might also be an good option to address AEr. As developer \#20 stated ``\textit{Consideration of people and organizational aspects of the architecture as well as technical aspects. Investment in people: training, study time, mentoring, etc}".

In addition, there are two less frequently mentioned measures: restarting a project or rewriting the architecture from scratch, and avoiding the systems growing larger than intended by controlling the size and functional diversity deliberately.


\section{Discussion}\label{sec:discussion}
\subsection{Interpretation of Results}\label{subsec:discussion}
\textbf{RQ1: Terms and manifestation of architecture erosion}. The results of RQ1.1 (see Table \ref{T:terms}) show that most of the developers prefer to use the term ``\textit{erode/erosion}” to describe the AEr phenomenon, followed by ``\textit{decay}" and other less-frequently used terms. Regarding the results of RQ1.2, we found that developers usually describe the phenomenon of AEr from four perspectives: structure, quality, maintenance, and evolution. All the four perspectives are worth investigating with further research; while there are some literature linking each perspective with AEr (see e.g., \cite{RN362} and \cite{RN436} for structure, \cite{RN329} for quality, and \cite{RN385} for maintenance and evolution), this investigation needs to be more systematic.

\textbf{RQ2: Causes and consequences of architecture erosion}. We identified 15 causes of AEr (see Table \ref{T:causes}). Inappropriate architecture changes is the most frequently mentioned reason that leads to AEr; this aligns with recent studies (e.g., \cite{RN353}) on architectural changes. 
Table \ref{T:causes} reveals that, alongside technical factors, non-technical factors are not trivial. In fact, the non-technical aspects seem to reinforce each other; for example, ``lack of communication" might induce the ``disconnection between architects and developers" and ``knowledge vaporization". The findings corroborate the results in \cite{RN221}, that good culture of communication and improving management skills are also important for architectural sustainability. 

According to Table \ref{T:consequences}, the potential consequences from AEr mirror the four perspectives mentioned in RQ1, and extend them with further effects. Specifically, in addition to the impact on maintenance and evolution, as well as run-time qualities (that match the perspectives of quality, maintenance and evolution), we also found the impact of AEr on development speed, cost of refactoring, and high staff turnover of developers. 
The findings show the significance of preventing and controlling AEr, and warn that massive cost might be invested in the degraded projects for tackling AEr.

\textbf{RQ3: Practices and tools for detecting architecture erosion}. Developers often employ practices and tools to indirectly detect AEr by indicators or symptoms (e.g., cyclic dependencies, architecture violations). Such practices and tools contribute to the identification of architecture issues that are reported in the literature (e.g., \cite{RN362}) and have a well-established connection to the phenomenon of AEr. Although they do not detect AEr per se, the findings suggest that there is a clear need for the software architecture community to devise dedicated tools on AEr detection.

Regarding the practices presented in Section \ref{subsec:RQ3}, our findings suggest that a trend analysis from the quality and evolution perspectives (e.g., architecture monitoring, checking the change of architectural smells) may help the development team better understand the health status of systems. For example, Merkle \cite{RN296} found that keeping track of an evolving architecture (especially the changes and trends) can contribute to stopping AEr by using tools like Structure101. 

\textbf{RQ4: Measures taken for controlling architecture erosion}. The results indicate that the measures can potentially help to alleviate the impact of AEr and prevent AEr during development. While there are few studies that validate such measures (e.g., \cite{RN398, RN325}), further evidence is required to attest to their merit as well as potential pitfalls. This evidence is important for convincing management to allocate resources to control AEr, such as conducting architecture evaluation and periodic maintenance, or architecture simplification. Otherwise, tackling AEr may not get priority over the urgent implementation of features and bug fixing. 
Consequently, developers cannot ignore and do nothing about the appearance and accumulation of system anomalies, and regular inspection and maintenance are indispensable for preventing and tackling AEr. In general, these measures can provide practitioners further guidance regarding architecture management on how to deal with AEr during architecture maintenance.

\subsection{Implications for Researchers and Practitioners}
\textbf{Terms of AEr}: Regarding the terms used for describing the AEr phenomenon, 
although it is difficult to establish a unified term to be used universally, we do recommend that researchers should define the used terms when they refer to AEr, in order to minimize the ambiguities and misunderstandings. Additionally, practitioners are also encouraged to find a common ground on understanding the AEr phenomenon for diminishing the ambiguity of the AEr concept and terms.

\textbf{Four perspectives of AEr}: It indicates that AEr manifests through structural issues, but mostly causes problems when it affects both run-time qualities (e.g., performance or reliability) and design-time qualities (e.g., maintainability and evolvability). Given AEr is a multifaceted phenomenon from the developers' perspective, and researchers can conduct more empirical studies to further investigate the characteristics of AEr. Additionally, the four perspectives of AEr should receive more attention from practitioners in their daily development activities, and these perspectives can be regarded as indicators for perceiving AEr. 

\textbf{Awareness of AEr}: 
The findings can also raise awareness among practitioners about potential causes of AEr that are not intuitive, such as the adoption of agile development. We would thus urge practitioners to pay more attentions to the grave consequences of AEr in order to request action at the management level. It is more likely that management will take measures or give a high priority to address AEr, when the aforementioned risks become explicit.

\textbf{Guidance for software development}: Having a good culture of communication and improving management skills (especially about the training and education of developers) are quite significant to developers for understanding the system design and structure. The findings can provide clues for practitioners to reduce the risk of AEr in their design and maintenance activities. Specifically, practitioners need to pay particular attention to the integrity of architecture when architecture changes happen.

\textbf{Approaches and tools with empirical evidence}: There is a need for researchers and practitioners to devise dedicated approaches and tools to detect and address AEr. Moreover, researchers can use the practices and tools reported in this study, to explore the scope, characteristics, and metrics of AEr, in order to provide a solid foundation on those tools. Furthermore, we encourage researchers and practitioners to explore the benefits and limitations by applying the approaches, tools, and measures for addressing AEr in practice.


\section{Threats to Validity}\label{sec:Threats}
The threats to the validity of this study are discussed by following the guidelines proposed by Wohlin \textit{et al}. \cite{RN265}. Internal validity is not considered, since this study does not address any causal relationships between variables.

\textbf{Construct validity} concerns if the theoretical and conceptual constructs are correctly interpreted and measured. In this study, there are two key threats. The first one concerns the search process related to data collection of AEr. To mitigate this threat, we leveraged Google to collect recommended popular online developer communities as many as possible, and excluded duplicate results and searched platforms with qualifiers (e.g., popular ``web/Android/PHP" communities, best developer communities ``in India"). Besides, we reviewed papers of AEr and summarized the most frequently-used terms about AEr in the literature (see Section \ref{subsec:terms}); these terms were used to derive our search string. The second threat lies in the process of manually extracting and analyzing the collected data. To partially mitigate this threat, we did a pilot execution of data filtering, extraction, and coding by the first and second authors, for reaching an agreement about all the terms used.

\textbf{External validity} concerns the extent to which we can generalize the research findings. A relevant threat concerns the representativeness of the selected online developer communities. To reduce this threat, we conducted a comprehensive analysis and selected several top-recommended online developer communities (see Section \ref{subsec:Research Process}), including the largest and most widely used Q\&A community by developers around the world (e.g., Stack Overflow) and other developer forums.

\textbf{Reliability} refers to the replicability of a study for generating the same or similar results. To alleviate this threat, we specified the process of our study design in a research protocol that can be used to replicate this work; Section \ref{sec:Study Design} presents the details of the study design, while the complete information and all instruments and data are available on the replication package \cite{RN288}. Moreover, pilot studies of data collection, filtering, and survey were conducted to mitigate misinterpretations and biases. Before the formal data analysis, we did a pilot data filtering, extraction, and coding by the first two authors. To eliminate personal biases, any conflicts and disagreements were discussed until an agreement was reached. Finally, we obtained a Cohen’s Kappa coefficient of 0.728 on the filtering process, which partially reduces this threat. 

\section{Conclusions and Future Work}\label{sec:Conclusions}
We conducted an empirical study to explores how developers perceive and discuss the phenomenon of AEr by collecting relevant information of AEr from the perspective of practitioners using three data sources (i.e., communities, surveys, interviews). The findings can provide practitioners with concrete measures to detect and control AEr, and provide researchers with the challenges on AEr.

We found that most developers described the phenomenon of AEr with terms like ``\textit{erode/erosion}”, ``\textit{decay}”, and ``\textit{degrade/degradation}”. When thinking about AEr, developers consider structural issues, but also the effect on run-time qualities, maintenance and evolution. Furthermore, besides the technical factors, non-technical factors play a big role in causing AEr. 
Despite the lack of dedicated tools for detecting AEr per se, developers employed associated practices and tools to detect the symptoms of AEr. To some extent, the practices and tools can help practitioners understand architectural structure and identify the eroding tendency of an architecture. Moreover, the identified measures can be employed during architecture implementation for effectively addressing AEr.

In the next step, we plan to detect and prevent AEr (semi-)automatically by establishing a dataset about symptoms of AEr and quantifying the degree of AEr from development artifacts (e.g., components). 

\balance
\bibliographystyle{ieeetr}
\bibliography{ref}

\begin{thebibliography}{10}

\bibitem{RN211}
L.~Bass, P.~Clements, and R.~Kazman, {\em Software Architecture in Practice
  (3rd Edition)}.
\newblock Addison-Wesley Professional, 3rd~ed., 2012.

\bibitem{RN222}
D.~E. Perry and A.~L. Wolf, ``Foundations for the study of software
  architecture,'' {\em ACM SIGSOFT Software Engineering Notes}, vol.~17, no.~4,
  pp.~40--52, 1992.

\bibitem{RN221}
L.~De~Silva and D.~Balasubramaniam, ``Controlling software architecture
  erosion: A survey,'' {\em Journal of Systems and Software}, vol.~85, no.~1,
  pp.~132--151, 2012.

\bibitem{RN210}
D.~M. Le, C.~Carrillo, R.~Capilla, and N.~Medvidovic, ``Relating architectural
  decay and sustainability of software systems,'' in {\em Proceedings of the
  13th Working IEEE/IFIP Conference on Software Architecture (WICSA)},
  pp.~178--181, IEEE, 2016.

\bibitem{RN213}
D.~M. Le, D.~Link, A.~Shahbazian, and N.~Medvidovic, ``An empirical study of
  architectural decay in open-source software,'' in {\em Proceedings of the
  IEEE International Conference on Software Architecture (ICSA)}, pp.~176--185,
  IEEE, 2018.

\bibitem{RN214}
Z.~Li and J.~Long, ``A case study of measuring degeneration of software
  architectures from a defect perspective,'' in {\em Proceedings of the 18th
  Asia-Pacific Software Engineering Conference (APSEC)}, pp.~242--249, IEEE,
  2011.

\bibitem{RN215}
I.~Macia, R.~Arcoverde, A.~Garcia, C.~Chavez, and A.~von Staa, ``On the
  relevance of code anomalies for identifying architecture degradation
  symptoms,'' in {\em Proceedings of the 16th European Conference on Software
  Maintenance and Reengineering (CSMR)}, pp.~277--286, IEEE, 2012.

\bibitem{RN212}
J.~Van~Gurp and J.~Bosch, ``Design erosion: problems and causes,'' {\em Journal
  of Systems and Software}, vol.~61, no.~2, pp.~105--119, 2002.

\bibitem{RN408}
F.~Brosig, P.~Meier, S.~Becker, A.~Koziolek, H.~Koziolek, and S.~Kounev,
  ``Quantitative evaluation of model-driven performance analysis and simulation
  of component-based architectures,'' {\em IEEE Transactions on Software
  Engineering}, vol.~41, no.~2, pp.~157--175, 2014.

\bibitem{RN409}
H.~P. Breivold and I.~Crnkovic, ``A systematic review on architecting for
  software evolvability,'' in {\em Proceedings of the 21st Australian Software
  Engineering Conference (ASWEC)}, pp.~13--22, IEEE, 2010.

\bibitem{RN364}
V.~V.~G. Neto, W.~Manzano, L.~Garcés, M.~Guessi, B.~Oliveira, T.~Volpato, and
  E.~Y. Nakagawa, ``Back-sos: towards a model-based approach to address
  architectural drift in systems-of-systems,'' in {\em Proceedings of the 33rd
  Annual ACM Symposium on Applied Computing (SAC)}, pp.~1461--1463, ACM, 2018.

\bibitem{RN258}
A.~Tahir, A.~Yamashita, S.~Licorish, J.~Dietrich, and S.~Counsell, ``Can you
  tell me if it smells?: A study on how developers discuss code smells and
  anti-patterns in stack overflow,'' in {\em Proceedings of the 22nd
  International Conference on Evaluation and Assessment in Software Engineering
  (EASE)}, pp.~68--78, ACM, 2018.

\bibitem{RN171}
F.~Tian, P.~Liang, and M.~A. Babar, ``How developers discuss architecture
  smells? an exploratory study on stack overflow,'' in {\em Proceedings of the
  IEEE International Conference on Software Architecture (ICSA)}, pp.~91--100,
  IEEE, 2019.

\bibitem{RN406}
V.~S. Sinha, S.~Mani, and M.~Gupta, ``Exploring activeness of users in {QA}
  forums,'' in {\em Proceedings of the 10th Working Conference on Mining
  Software Repositories (MSR)}, pp.~77--80, IEEE, 2013.

\bibitem{RN433}
S.~Adolph, W.~Hall, and P.~Kruchten, ``Using grounded theory to study the
  experience of software development,'' {\em Empirical Software Engineering},
  vol.~16, no.~4, pp.~487--513, 2011.

\bibitem{RN302}
R.~Mo, J.~Garcia, Y.~Cai, and N.~Medvidovic, ``Mapping architectural decay
  instances to dependency models,'' in {\em Proceedings of the 4th
  International Workshop on Managing Technical Debt (MTD)}, pp.~39--46, IEEE,
  2013.

\bibitem{RN271}
R.~Land, ``Software deterioration and maintainability–a model proposal,'' in
  {\em Proceedings of the 2nd Conference on Software Engineering Research and
  Practice in Sweden (SERPS)}, Citeseer, 2002.

\bibitem{RN252}
L.~Hochstein and M.~Lindvall, ``Combating architectural degeneration: a
  survey,'' {\em Information and Software Technology}, vol.~47, no.~10,
  pp.~643--656, 2005.

\bibitem{RN272}
A.~Bianchi, D.~Caivano, F.~Lanubile, and G.~Visaggio, ``Evaluating software
  degradation through entropy,'' in {\em Proceedings of the 7th International
  Software Metrics Symposium (METRICS)}, pp.~210--219, IEEE, 2001.

\bibitem{RN329}
T.~Wang, D.~Wang, and B.~Li, ``A multilevel analysis method for architecture
  erosion,'' in {\em Proceedings of the 31st International Conference on
  Software Engineering and Knowledge Engineering (SEKE)}, pp.~443--566, KSI,
  2019.

\bibitem{RN273}
M.~Soliman, A.~R. Salama, M.~Galster, O.~Zimmermann, and M.~Riebisch,
  ``Improving the search for architecture knowledge in online developer
  communities,'' in {\em Proceedings of the IEEE International Conference on
  Software Architecture (ICSA)}, pp.~186--195, IEEE, 2018.

\bibitem{RN274}
T.~Bi, P.~Liang, and A.~Tang, ``Architecture patterns, quality attributes, and
  design contexts: How developers design with them,'' in {\em Proceedings of
  the 25th Asia-Pacific Software Engineering Conference (APSEC)}, pp.~49--58,
  IEEE, 2018.

\bibitem{RN278}
A.~Pal, S.~Chang, and J.~A. Konstan, ``Evolution of experts in question
  answering communities,'' in {\em Proceedings of the 6th International
  Conference on Weblogs and Social Media (ICWSM)}, pp.~274--281, AAAI, 2012.

\bibitem{RN279}
S.~Wang, D.~Lo, and L.~Jiang, ``An empirical study on developer interactions in
  stackoverflow,'' in {\em Proceedings of the 28th Annual ACM Symposium on
  Applied Computing (SAC)}, pp.~1019--1024, ACM, 2013.

\bibitem{RN280}
J.~Tsay, L.~Dabbish, and J.~Herbsleb, ``Influence of social and technical
  factors for evaluating contribution in github,'' in {\em Proceedings of the
  36th International Conference on Software Engineering (ICSE)}, pp.~356--366,
  ACM, 2014.

\bibitem{RN50}
P.~Runeson and M.~Höst, ``Guidelines for conducting and reporting case study
  research in software engineering,'' {\em Empirical Software Engineering},
  vol.~14, no.~2, pp.~131--164, 2009.

\bibitem{RN117}
V.~R. Basili, G.~Caldiera, and H.~D. Rombach, ``The goal question metric
  approach,'' {\em Encyclopedia of Software Engineering}, pp.~528--532, 1994.

\bibitem{RN245}
L.~Zhang, Y.~Sun, H.~Song, F.~Chauvel, and H.~Mei, ``Detecting architecture
  erosion by design decision of architectural pattern,'' in {\em Proceedings of
  the 23rd International Conference on Software Engineering and Knowledge
  Engineering (SEKE)}, pp.~758--763, KSI, 2011.

\bibitem{RN392}
J.~Adersberger and M.~Philippsen, ``Reflexml: Uml-based architecture-to-code
  traceability and consistency checking,'' in {\em Proceedings of the 5th
  European Conference on Software Architecture (ECSA)}, vol.~6903,
  pp.~344--359, Springer, 2011.

\bibitem{RN260}
J.~Vassileva, ``Toward social learning environments,'' {\em IEEE Transactions
  on Learning Technologies}, vol.~1, no.~4, pp.~199--214, 2008.

\bibitem{RN288}
R.~Li, P.~Liang, M.~Soliman, and P.~Avgeriou, ``Replication package for the
  paper: Understanding architecture erosion: The practitioners' perceptive,''
  \url{https://doi.org/10.5281/zenodo.4481564}, 2021.

\bibitem{RN569}
M.~Soliman, M.~Galster, A.~R. Salama, and M.~Riebisch, ``Architectural
  knowledge for technology decisions in developer communities: An exploratory
  study with stackoverflow,'' in {\em Proceedings of the 13th Working IEEE/IFIP
  Conference on Software Architecture (WICSA)}, pp.~128--133, IEEE, 2016.

\bibitem{RN261}
J.~Cohen, ``A coefficient of agreement for nominal scales,'' {\em Educational
  and Psychological Measurement}, vol.~20, no.~1, pp.~37--46, 1960.

\bibitem{RN116}
F.~Shull, J.~Singer, and D.~I. Sjøberg, {\em Guide to advanced empirical
  software engineering}.
\newblock Springer, 2007.

\bibitem{RN405}
S.~E. Hove and B.~Anda, ``Experiences from conducting semi-structured
  interviews in empirical software engineering research,'' in {\em Proceedings
  of the 11th IEEE International Software Metrics Symposium (METRICS)},
  pp.~10--23, IEEE, 2005.

\bibitem{RN112}
``{MAXQDA},'' \url{https://www.maxqda.com/}.
\newblock accessed on April 30, 2020.

\bibitem{RN404}
Z.~Li, P.~Avgeriou, and P.~Liang, ``A systematic mapping study on technical
  debt and its management,'' {\em Journal of Systems and Software}, vol.~101,
  pp.~193--220, 2015.

\bibitem{RN266}
D.~Sturtevant, ``Modular architectures make you agile in the long run,'' {\em
  IEEE Software}, vol.~35, no.~1, pp.~104--108, 2017.

\bibitem{RN421}
B.~Foote and J.~Yoder, ``Big ball of mud,'' {\em Pattern Languages of Program
  Design}, vol.~4, pp.~654--692, 1997.

\bibitem{RN309}
N.~Kumar, ``Software architecture validation methods, tools support and case
  studies,'' in {\em Emerging Research in Computing, Information, Communication
  and Applications}, chapter~32, pp.~335--345, Springer, 2016.

\bibitem{RN310}
R.~S. Sangwan, P.~Vercellone-Smith, and P.~A. Laplante, ``Structural epochs in
  the complexity of software over time,'' {\em IEEE Software}, vol.~25, no.~4,
  pp.~66--73, 2008.

\bibitem{RN313}
M.~K. Gopal, ``Design quality metrics on the package maintainability and
  reliability of open source software,'' {\em International Journal of
  Intelligent Engineering and Systems}, vol.~9, no.~4, pp.~195--204, 2016.

\bibitem{RN301}
N.~Sangal, E.~Jordan, V.~Sinha, and D.~Jackson, ``Using dependency models to
  manage complex software architecture,'' in {\em Proceedings of the 20th
  Annual ACM SIGPLAN Conference on Object-oriented Programming, Systems,
  Languages, and Applications (OOPSLA)}, pp.~167--176, ACM, 2005.

\bibitem{RN303}
R.~L. Nord, I.~Ozkaya, P.~Kruchten, and M.~Gonzalez-Rojas, ``In search of a
  metric for managing architectural technical debt,'' in {\em Proceedings of
  the 2012 Joint Working IEEE/IFIP Conference on Software Architecture and
  European Conference on Software Architecture (WICSA/ECSA)}, pp.~91--100,
  IEEE, 2012.

\bibitem{RN305}
A.~Mokni, C.~Urtado, S.~Vauttier, M.~Huchard, and H.~Y. Zhang, ``A formal
  approach for managing component-based architecture evolution,'' {\em Science
  of Computer Programming}, vol.~127, pp.~24--49, 2016.

\bibitem{RN306}
A.~Mokni, M.~Huchard, C.~Urtado, S.~Vauttier, and Y.~Zhang, ``An evolution
  management model for multi-level component-based software architectures,'' in
  {\em Proceedings of the 27th International Conference on Software Engineering
  and Knowledge Engineering (SEKE)}, pp.~674--679, KSI, 2015.

\bibitem{RN415}
G.~C. Murphy, D.~Notkin, and K.~Sullivan, ``Software reflexion models: Bridging
  the gap between source and high-level models,'' in {\em Proceedings of the
  3rd ACM SIGSOFT Symposium on Foundations of Software Engineering (FSE)},
  pp.~18--28, ACM, 1995.

\bibitem{RN264}
M.~D’Ambros, H.~Gall, M.~Lanza, and M.~Pinzger, ``Analysing software
  repositories to understand software evolution,'' in {\em Software Evolution},
  chapter~3, pp.~37--67, Springer, 2008.

\bibitem{RN445}
T.~Sharma, P.~Singh, and D.~Spinellis, ``An empirical investigation on the
  relationship between design and architecture smells,'' {\em Empirical
  Software Engineering}, vol.~25, no.~5, pp.~4020--4068, 2020.

\bibitem{RN416}
M.~Shahin, P.~Liang, and M.~A. Babar, ``A systematic review of software
  architecture visualization techniques,'' {\em Journal of Systems and
  Software}, vol.~94, pp.~161--185, 2014.

\bibitem{RN431}
M.~W. Godfrey and E.~H. Lee, ``Secrets from the monster: Extracting mozilla’s
  software architecture,'' in {\em Proceedings of the 2nd International
  Symposium on Constructing Software Engineering Tools (CoSET)}, pp.~15--23,
  Citeseer, 2000.

\bibitem{RN362}
I.~Macia, J.~Garcia, D.~Popescu, A.~Garcia, N.~Medvidovic, and A.~von Staa,
  ``Are automatically-detected code anomalies relevant to architectural
  modularity?: An exploratory analysis of evolving systems,'' in {\em
  Proceedings of the 11th Annual International Conference on Aspect-oriented
  Software Development (AOSD)}, pp.~167--178, ACM, 2012.

\bibitem{RN436}
C.~B. Jaktman, J.~Leaney, and M.~Liu, ``Structural analysis of the software
  architecture—a maintenance assessment case study,'' in {\em Proceedings of
  the 1st Working IEEE/IFIP Conference on Software Architecture (WICSA)},
  pp.~455--470, Springer, 1999.

\bibitem{RN385}
J.~Brunet, R.~A. Bittencourt, D.~Serey, and J.~Figueiredo, ``On the
  evolutionary nature of architectural violations,'' in {\em Proceedings of the
  19th Working Conference on Reverse Engineering (WCRE)}, pp.~257--266, IEEE,
  2012.

\bibitem{RN353}
D.~M. Le, P.~Behnamghader, J.~Garcia, D.~Link, A.~Shahbazian, and
  N.~Medvidovic, ``An empirical study of architectural change in open-source
  software systems,'' in {\em Proceedings of the 12th IEEE/ACM Working
  Conference on Mining Software Repositories (MSR)}, pp.~235--245, IEEE, 2015.

\bibitem{RN296}
B.~Merkle, ``Stop the software architecture erosion: building better software
  systems,'' in {\em Proceedings of the 25th Annual ACM SIGPLAN Conference on
  Object-Oriented Programming, Systems, Languages, and Applications
  (SPLASH/OOPSLA) Companion}, pp.~129--138, ACM, 2010.

\bibitem{RN398}
S.~Gerdes, S.~Jasser, M.~Riebisch, S.~Schröder, M.~Soliman, and T.~Stehle,
  ``Towards the essentials of architecture documentation for avoiding
  architecture erosion,'' in {\em Proceedings of the 10th European Conference
  on Software Architecture Workshops (ECSA)}, pp.~1--4, ACM, 2016.

\bibitem{RN325}
M.~Stal, ``Refactoring software architectures,'' in {\em Agile Software
  Architecture}, chapter~3, pp.~63--82, Elsevier, 2014.

\bibitem{RN265}
C.~Wohlin, P.~Runeson, M.~Höst, M.~C. Ohlsson, B.~Regnell, and A.~Wesslén,
  {\em Experimentation in Software Engineering}.
\newblock Springer Science \& Business Media, 2012.

\end{thebibliography}
\end{document}